# Reversible modulation of metal-insulator transition in VO$_2$ via chemically-induced oxygen migration

Kun Han[‡ a,b], Hanyu Wang[‡c+], Liang Wu[*d,e], Yu Cao[f], Dong-Chen Qi[g], Changjian Li[h], Zhen Huang[a], Xiao Li[*c] and X. Renshaw Wang[*b,i]

[a] Information Materials and Intelligent Sensing Laboratory of Anhui Province, Institutes of Physical Science and Information Technology, Anhui University, Hefei 230601, China

[b] Division of Physics and Applied Physics, School of Physical and Mathematical Sciences, Nanyang Technological University, 21 Nanyang Link, 637371, Singapore

[c] Center for Quantum Transport and Thermal Energy Science (CQTES), School of Physics and Technology, Nanjing Normal University, Nanjing 210023, China

[d] Faculty of Material Science and Engineering, Kunming University of Science and Technology, Kunming, Yunnan 650093, China

[e] Foshan (Southern China) Institute for New Materials, Foshan, 528247, China

[f] Department of Electrical and Computer Engineering, National University of Singapore, 4 Engineering Drive 3, 117583, Singapore

[g] Centre for Materials Science, School of Chemistry and Physics, Queensland University of Technology, Brisbane, Queensland 4001, Australia

[h] Department of Materials Science and Engineering, Southern University of Science and Technology, 518055 Shenzhen, Guangdong, China

[i] School of Electrical and Electronic Engineering, Nanyang Technological University, 50 Nanyang Ave, 639798, Singapore

[‡] These authors contributed equally to this work.
[+]Current address: 1. Key Laboratory of Materials Physics, Institute of Solid State Physics, HFIPS, Chinese Academy of Science, Hefei 230031, China; 2. Science Island Branch of the Graduate School, University of Science and Technology of China, Hefei 230026, China.
* Correspondence to: liangwu@kust.edu.cn; lixiao@njnu.edu.cn; renshaw@ntu.edu.sg

**ABSTRACT**

Metal-insulator transitions (MIT), an intriguing correlated phenomenon induced by the subtle competition of the electrons' repulsive Coulomb interaction and kinetic energy, is of great potential use for electronic applications due to the dramatic change in resistivity. Here, we demonstrate a reversible control of MIT in $VO_2$ films via oxygen stoichiometry engineering. By facilely depositing and dissolving a water-soluble yet oxygen-active $Sr_3Al_2O_6$ capping layer atop the $VO_2$ at room temperature, oxygen ions can reversibly migrate between $VO_2$ and $Sr_3Al_2O_6$, resulting in a gradual suppression and a complete recovery of MIT in $VO_2$. The migration of the oxygen ions is evidenced in a combination of transport measurement, structural characterization and first-principles calculations. This approach of chemically-induced oxygen migration using a water-dissolvable adjacent layer could be useful for advanced electronic and iontronic devices and studying oxygen stoichiometry effects on the MIT.

Originated from the subtle and enigmatic competition of the electrons' repulsive Coulomb interaction and kinetic energy, the metal-insulator transition (MIT) shows a drastic change in resistivity, which has been observed in diverse systems, such as the Mott-Hubbard MIT[1,2], Peierls MIT[3], Anderson MIT[4,5], and spin-orbit coupling (SOC)-controlled MIT[6,7]. Besides its appealing fundamental aspect of the properties, its application in data processing and storage due to the fine tuning of resistivity is highly desirable for future low energy electronics[8–10]. One quintessential example is the archetypal MIT in $VO_2$, which shows up to 5 orders of magnitude resistivity change across the MIT temperature ($T_{MIT}$) of ~341 K[10]. The properties of $VO_2$ certainly change from an insulating monoclinic [$P2_1/c$, namely $VO_2(M)$] phase (*M*-phase) to a metallic rutile [$P4_2/mmm$, namely $VO_2(R)$][9–11] (*R*-phase), the MIT can be effectively controlled with oxygen stoichiometry[12–14]. Hence, to widen the functionality and gain a deeper insight into the MIT of $VO_2$, reversibly metallizing and recovering the insulating state has been a long-standing goal[13,15].

To date, typical methods to manipulate the MIT are vacuum annealing[16], hydrogenation[17], ionic liquid gating[13,18]. Moreover, there are a large number of methods to modulate the MIT, such as chemical doping[19], strain engineering[20], defect engineering[21], and voltage-induced ionic migration[22], etc. Recently, the water-soluble $Sr_3Al_2O_6$ (SAO) attracts much attention because of its capability in synthesizing freestanding functional thin films[23–26]. Also, SAO can act as an oxygen pump to tune the oxygen stoichiometry of its adjacent oxide layers even in an amorphous form[27]. In this letter, we report a facile and reliable method to reversibly modulate the MIT in epitaxial $VO_2$ thin-film readily by depositing and dissolving an amorphous-SAO (*a*-SAO) capping layer. Specifically, the MIT in $VO_2$ thin film is gradually suppressed or even eliminated by increasing the thickness of the capping layer and almost fully recovered after dissolving the capping layer in water.

**Fig. 1a** shows the temperature dependence of the normalized resistivity $R(T)/R(400\text{ K})$ of 5 nm $VO_2$ grown on $Al_2O_3$ (0001) substrates, denoted as $VO_2//Al_2O_3$ (0001) in this study, with the thicknesses of the *a*-SAO capping layer varying from 0 to 100 nm. Both VO$_2$ and SAO were deposited by pulsed laser deposition (PLD) using KrF excimer laser ($\lambda$ = 248 nm) with a laser fluence energy density of ~2 J cm$^{-2}$ and a repetition rate of 5 Hz. The substrate temperature and oxygen partial pressure were fixed at 500 °C and 1 mTorr, respectively. The epitaxial VO$_2$ thin films were grown on Al$_2$O$_3$ (0001) substrates. After deposition, the samples were post-annealed in an oxygen partial pressure of 5 mTorr at 500 °C for 1 h and then cooled down to room temperature at a constant rate of 10 °C/min in the same oxygen partial pressure[28,29]. The *a*-SAO capping layers were also fabricated by PLD at room temperature and an oxygen partial pressure of 5 mTorr. Commercial Vanadium single crystal (100)-orientated metal target with 99.999% purity (from Goodfellow) and a sintered SAO polycrystalline pallet were used as targets[23,27].

A single VO$_2$//Al$_2$O$_3$ (0001) sample with a size of 10 × 10 mm$^2$ was prepared and cut into 16 pieces of similar size. Then, SAO with different thicknesses was deposited onto the cut VO$_2$//Al$_2$O$_3$ (0001) pieces. The pristine state of VO$_2$ exhibited an MIT at around 340 K, accompanied by a thermal hysteresis, which is the characteristic of first-order phase transition[10]. A progressive suppression of the MIT and metallization of the VO$_2$ film were observed as the thickness of *a*-SAO capping layer ($t_{cap}$) increased until the MIT is nearly eliminated for $t_{cap} \geq 50$ nm. This deterioration of MIT was believed to result from a redox-induced oxygen migration out of VO$_2$, *i.e.*, vacancy formation in VO$_2$.[12,30] The reduction of vanadium valence state from 4+ to 3+ due to the electron doping via the oxygen vacancies are responsible for the metallization of the oxygen-deficient $VO_2$ down to lower temperatures.[14,31] Additionally, the MITs of 10, 12.5 and 15 nm thick $VO_2$ films can only be partially suppressed by capping an *a*-SAO layer (**Fig. S1**). This suggests that the oxygen vacancies in $VO_2$ are highly mobile at room temperature and can easily migrate beyond 15 nm, rather than highly concentrated within the top several unit cells of the surface region. Because the $Al_2O_3$ substrate is not capable of accommodating the oxygen vacancy, when the thickness of the $VO_2$ is relatively small, *i.e.*, 5 nm, a large amount of oxygen vacancies may accumulate and be refilled within the whole $VO_2$ films. When the $VO_2$ is thick, *i.e.*, 80 nm, the MIT of the 80-nm $VO_2$//$Al_2O_3$ (0001) is slightly suppressed even for $t_{cap}$ = 100 nm (**Fig. 1b,d**), which indicates the oxygen vacancies created by the *a*-SAO capping layer grown at room temperature migrates deep into the $VO_2$, resulting in diluted oxygen vacancies across the film and, consequently, a weaker insulating state. It is well understood that both the amount and distribution of oxygen vacancies are highly temperature-dependent[32]. To this end, a series of growth-temperature-dependent 100-nm SAO/80-nm $VO_2$//$Al_2O_3$ (0001) samples were prepared, because the oxygen diffusion coefficient significantly increases at elevated temperatures[33]. **Fig. 1c** shows that the MIT of 80-nm $VO_2$//$Al_2O_3$ (0001) is gradually suppressed with increasing the growth temperature of the SAO capping layer, which validates that more oxygen vacancies are created and migrate deep into the thick $VO_2$ films at higher temperatures. The oxygen deprivation of $VO_2$ by the SAO capping layer is schematically shown in the inset of **Fig. 1e**.

Previous studies have shown that the oxygen vacancies can be removed at the surfaces of many transition metal oxides, such as $SrTiO_3$[27,34,35], $TiO_2$[36,37] and $BaTiO_3$[34,38], at room temperature when exposed to water. Here, we show that MIT can be nearly restored in the $VO_2$//$Al_2O_3$ (0001) by dissolving the *a*-SAO capping layer by de-ionized (DI) water. **Fig. 2a** shows the temperature dependence of the normalized resistivity $R(T)/R(400\ K)$ for 5-nm $VO_2$//$Al_2O_3$ (0001) before and after dissolving the *a*-SAO capping layer. Noted that the effect of water on the transport property of a pristine $VO_2$ thin film is negligible, because no obvious change was found after dipping the bare 5-nm $VO_2$//$Al_2O_3$ (0001) in water for 48 h. By contrast, the suppressed MIT in 5-nm $VO_2$ thin film was fully recovered to its pristine state right after dissolving the 1-nm *a*-SAO capping layer. For a thicker capping layer, the MIT was first partially restored just after dissolving the 20 nm capping layer, then almost fully restored after dipping in water for 48 h. These phenomena can be well understood by taking the concentration and distribution of oxygen vacancies created during the deposition of *a*-SAO capping layers. To be specific, a thicker *a*-SAO capping layer can induce more oxygen vacancies in $VO_2$ thin films. Thus, longer water treatment was needed to completely incorporate the oxygen vacancies to the initial insulating state of *M*-phase $VO_2$.

On the other hand, the 80-nm $VO_2$//$Al_2O_3$ (0001) samples with a 100-nm SAO capping layer grown at different temperatures show different resistivity changes ($R(T)/R(400\ K)$) (**Fig. 2b and Fig. S2**). As mentioned above, the higher the growth temperature of the SAO capping layer was used, the more oxygen vacancies were created, consequently, the deeper these oxygen vacancies penetrated into the $VO_2$ film. We also noticed that the resistivity of 80-nm $VO_2$ with the SAO layer grown at room temperature almost returns to the pristine state after dissolving the capping layer, while the drastically suppressed MIT doesn't restore after dissolving the 500 °C-deposited SAO capping layer by dipping in DI water for 48 h. These experiments prove that the high growth temperature facilitates the creation and migration of the oxygen vacancies, leading to a $VO_2$ film with thicker and denser oxygen vacancies. Consistently, we found that the suppression of MIT in the 80-nm $VO_2$ returned to the pristine state by annealing in 5 mTorr oxygen partial pressure at 500 °C for 1 h, which is consistent with the previous study[39]. Therefore, the MIT in both thin and thick $VO_2$ films can be reversibly controlled. It is worth noting that, the diffusion of Al or Sr from SAO into $VO_2$ and damage of $VO_2$ during the high-energy PLD process may also suppress the MIT of $VO_2$, however, which would induce an irreversible effect on the MIT of $VO_2$.

To gain a deeper understanding of the relation between the phase transition of $VO_2$ and the thickness of the *a*-SAO capping layer, Raman spectra and X-ray diffraction (XRD) (**Fig. 3**) were measured at room temperature. Raman spectra are sensitive to $VO_2$ *M*-phase, whose characteristic peaks are located at 194, 223, and 613 $cm^{-1}$. The intensity of the Raman characteristic peaks can represent the relative amount of the *M*-phase $VO_2$. **Fig. 3a** shows that the characteristic peaks of the 10-nm $VO_2$ film gradually decreased and eventually disappeared as the thickness of the *a*-SAO capping layer increased, which indicates a gradual decrease of the amount of the insulating *M*-phase $VO_2$[40]. The Raman peaks of 20-nm $VO_2$ are partially suppressed even after capping a 100-nm *a*-SAO layer (**Fig. 3b**) and the suppressed Raman peaks return to their pristine states after dissolving the *a*-SAO capping layer. This is consistent with the XRD in **Fig. 3c**, as the 10-nm VO$_2$ (020) peak shifted to a smaller angle after capping *a*-SAO with expanded lattice constant induced by oxygen vacancies[13,18]. In addition, as elaborated above, the oxygen vacancies are highly mobile, the XRD of the 80 nm thick VO$_2$ shows the imperceptible difference (**Fig. 3d**).

Furthermore, we performed density functional theory calculations (detailed calculation information can be found in **Supplementary Information Section 4**) for structural and electronic properties of VO$_2$ in both monoclinic and tetragonal rutile phases to gain an intuitive insight into the suppression of MIT in VO$_2$. Given that the *a*-SAO capping layer experimentally creates oxygen vacancies in VO$_2$, the effects from the oxygen vacancy on the two phases of VO$_2$ are focused on in the calculation. **Fig. 4a-d** demonstrates the pristine geometric structures with the vertical directions along crystallographic (010) growth axis for two phases and corresponding ones after introducing one oxygen vacancy, where pristine superlattices include 64 VO$_2$ formula units and the introduction of one oxygen vacancy corresponds to a vacancy concentration of 1/128. When the oxygen vacancy is absent, the *M*-phase has parallel chains of V-V dimerization along [$\bar{1}$10] axis, with two alternating V-V bond lengths of 2.52 and 3.17 Å, due to the Peierls distortion[41,42]. In contrast, there are equivalent V-V bonds of 2.79 Å along [100] axis for the *R* structure. After introducing the oxygen vacancy shown by the red ball in **Fig. 4**, most of long- and short-bond lengths in the *M*-phase are changed within ~0.1 Å compared with the pristine structure, except for three neighbouring bonds, as denoted by $d_{1-3}$ and dashed lines in **Fig. 4b**, having large changes by -0.37, +0.78 and -0.37 Å. The V-V bond lengths in the *R* phase vary within the range of 2.52-3.35 Å and some of them in the neighbourhood of the oxygen vacancy also become dimerized with alternating bonds (dashed lines in **Fig. 4d**), structurally similar to the *M*-phase.

**Fig. 4e-h** shows electronic structures of $VO_2$ before and after creating the oxygen vacancy. For pristine $VO_2$, the *M*- and *R*-phases exhibit an insulating state with a band gap of 0.7 eV and a metallic state, respectively, according to their band structures and densities of states. With the oxygen vacancy, while the metallicity of the *R*-phase is well kept, the conduction band edge of the *M* phase is partially occupied, indicating electron doping from the oxygen vacancy, agreeing with previous results[43,44]. By the charge density analyses of occupied conduction states, besides electronic densities distributed around the oxygen vacancy, there is also a relatively delocalized contribution that demonstrating the character of itinerant electrons, as illustrated in **Fig. S5**. Moreover, hole doping is found to increase with the number of oxygen vacancies. With considerable oxygen vacancies, the resistivity of the *M*-phase may decrease by the vacancy migration and the itinerant carriers, compared with the pristine $VO_2$.

On the other hand, the energy differences between the two phases are computed to indicate the feasibility of structural phase transition[45]. For $VO_2$ both without and with the oxygen vacancy, the *M*-phase has lower energy than the *R*-phase, indicating the *M*-phase is more stable at zero temperature. The absolute energy difference decreases from 139 to 83 meV per V atom by introducing one oxygen vacancy, and consequently, the phase transition is likely to occur more feasibly, *i.e.*, at a lower temperature, which is consistent with our experimental results (**Fig. 1** and **Fig. 2**). Therefore, the variation of the resistivity can be associated with the phase transition between insulating *M*-phase and metallic *R*-phase, besides the hole-doping from the vacancy itself. Two phases are inclined to become similar in both structural and electronic properties with the help of the oxygen vacancy, leading to the suppression of MIT in VO$_2$.

In summary, we have demonstrated a facile method to reversibly control the MIT in VO$_2$ thin film by depositing and dissolving a water-soluble *a*-SAO capping layer. Both Raman and XRD measurements indicate the suppression of MIT in VO$_2$ is ascribed to redox-reaction-induced oxygen migration at the heterointerface between VO$_2$ thin film and *a*-SAO capping layer. The ability to reversible control the MIT in VO$_2$ on a large scale, just by depositing and dissolving an *a*-SAO capping layer at room temperature, provides a platform to study oxygen stoichiometry effects on the MIT and oxide electronic and iontronic devices.

**Acknowledgement**

X.R.W. acknowledges support from the Tier 2 (Grant No. MOE-T2EP50120-0006) and Tier 3 (Grant No. MOE2018-T3-1-002) from Singapore Ministry of Education, and the Singapore National Research Foundation (NRF) under the competitive Research Programs (CRP Grant No. NRF-CRP21-2018-0003), and the Agency for Science, Technology and Research (A*STAR) under its AME IRG grant (Project No.

A20E5c0094). L.W. acknowledges supports from Foshan (Southern China) Institute for New Materials (2021AYF25014). X.R.W. thanks Ke Huang, Ariando and T. Venky Ventakesan for their help. Z.H. acknowledges the support from the National Natural Science Foundation of China (Grant No. 12074001). X.L. acknowledges support from the National Natural Science Foundation of China (Grant No. 11904173) and the Jiangsu Specially-Appointed Professor. D. Q. acknowledges the support of the Australian Research Council (Grant No. FT160100207).

**Data availability**

The data that support the findings of this study are available from the corresponding author upon reasonable request.

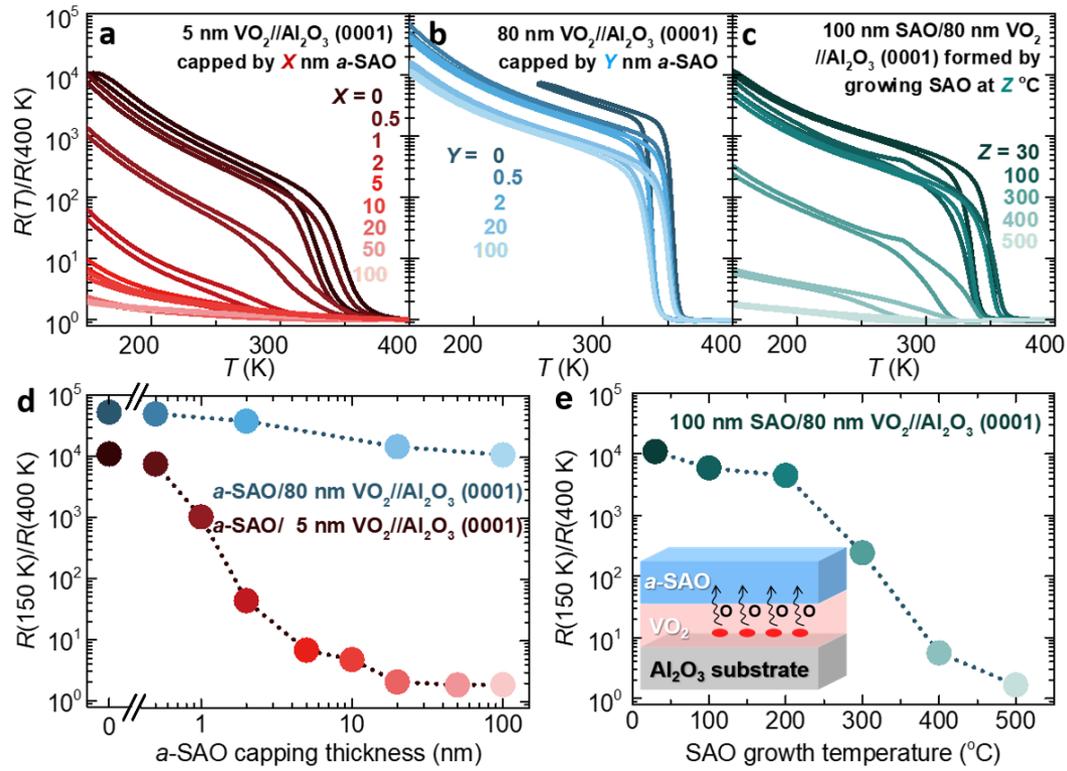

**Figure 1. Suppression of the MIT in epitaxial VO₂ thin films.** Normalized resistivity $R(T)/R(400\text{ K})$ versus temperature $T$ for various $a$-SAO capping layer thicknesses varying from 0 to 100 nm for (**a**) 5 nm VO₂//Al₂O₃ (0001) and (**b**) 80 nm VO₂//Al₂O₃ (0001). (**c**) Normalized resistivity $R(T)/R(400\text{ K})$ versus temperature, $T$, for 80 nm VO₂//Al₂O₃ (0001) with 100 nm SAO grown at different temperatures varying from room temperature to 500 °C. Original data can be found in **Fig. S3**. (**d**) Comparison of resistivity changes ($R(150\text{ K})/R(400\text{ K})$) as a function of $a$-SAO capping layer thicknesses for 5 and 80 nm thick VO₂ grown on Al₂O₃ (0001). (**e**) Resistivity changes ($R(150\text{ K})/R(400\text{ K})$) of 80 nm VO₂//Al₂O₃ (0001) with 100 nm SAO capping layers grown at different temperatures. The inset in (**e**) schematically shows the redox reaction at the SAO/VO₂ interface.

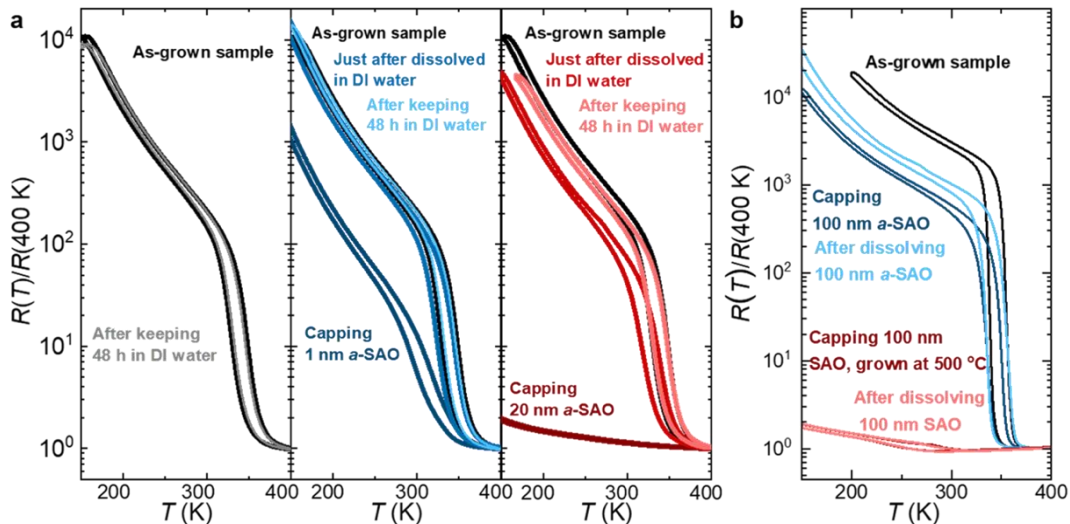

**Figure 2. Reversible MIT in epitaxial VO₂ thin films.** The as-grown samples refer to the 5-nm VO₂//Al₂O₃ (0001) prior to the deposition of the SAO layer. Comparison of normalized resistivity $R(T)/R(400\ K)$ versus temperature, $T$, for (**a**) 5-nm VO₂//Al₂O₃ (0001) capped with 1 and 20 nm thick $a$-SAO layers, just after the $a$-SAO dissolved in DI water, after dipping in DI water for 48 h. Original data can be found in **Fig. S4**. (**b**) Normalized resistivity ($R(T)/R(400\ K)$) of 80-nm VO₂//Al₂O₃ (0001) after capping with 100-nm SAO grown at room temperature and 500 °C, and dipping in DI water.

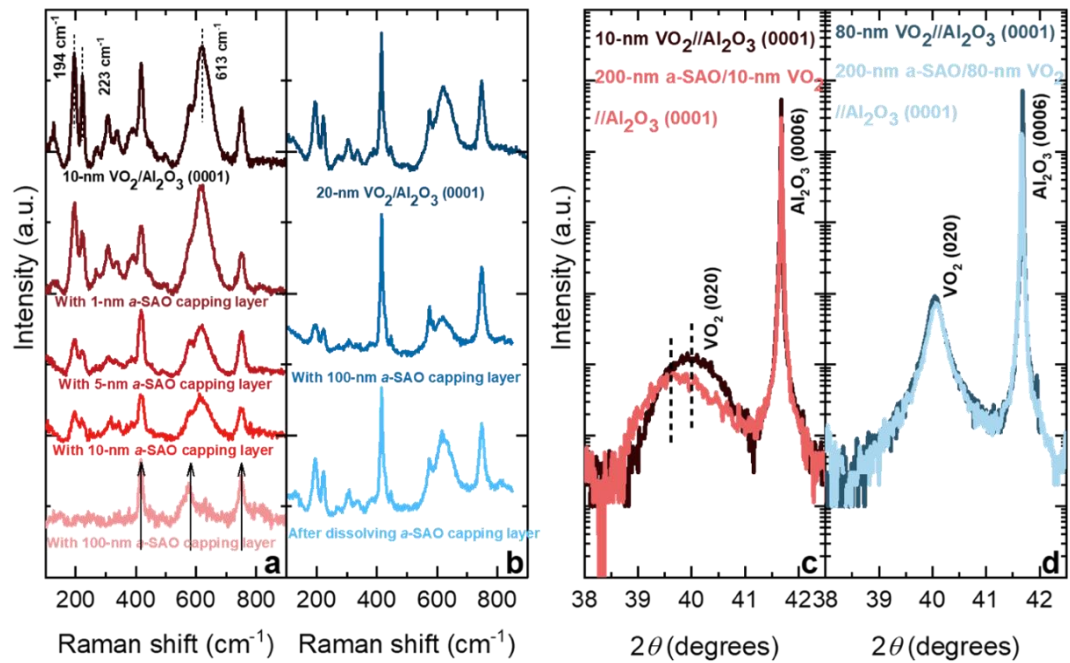

**Figure 3. Raman spectra and XRD measurement of epitaxial VO₂ thin films.** Raman spectra of (**a**) 10 nm VO₂//Al₂O₃ (0001) capped with different thicknesses of $a$-SAO layers, and (**b**) 20 nm VO₂//Al₂O₃ (0001) with a 100 nm $a$-SAO capping layer and after $a$-SAO was dissolved. Identified Raman peaks corresponding to the $M$ phase of VO₂. In particular $\omega_{V1}$ (194 cm⁻¹), $\omega_{V2}$ (223 cm⁻¹), and 613 cm⁻¹ peaks are marked with black dashed lines. The Raman peaks of Al₂O₃ (0001) substrate are marked with black arrows. $2\theta$-$\omega$ scans of (**c**) 10 nm VO₂//Al₂O₃ (0001) and (**d**) 80 nm VO₂//Al₂O₃ (0001) with and without a 200 nm $a$-SAO capping layer.

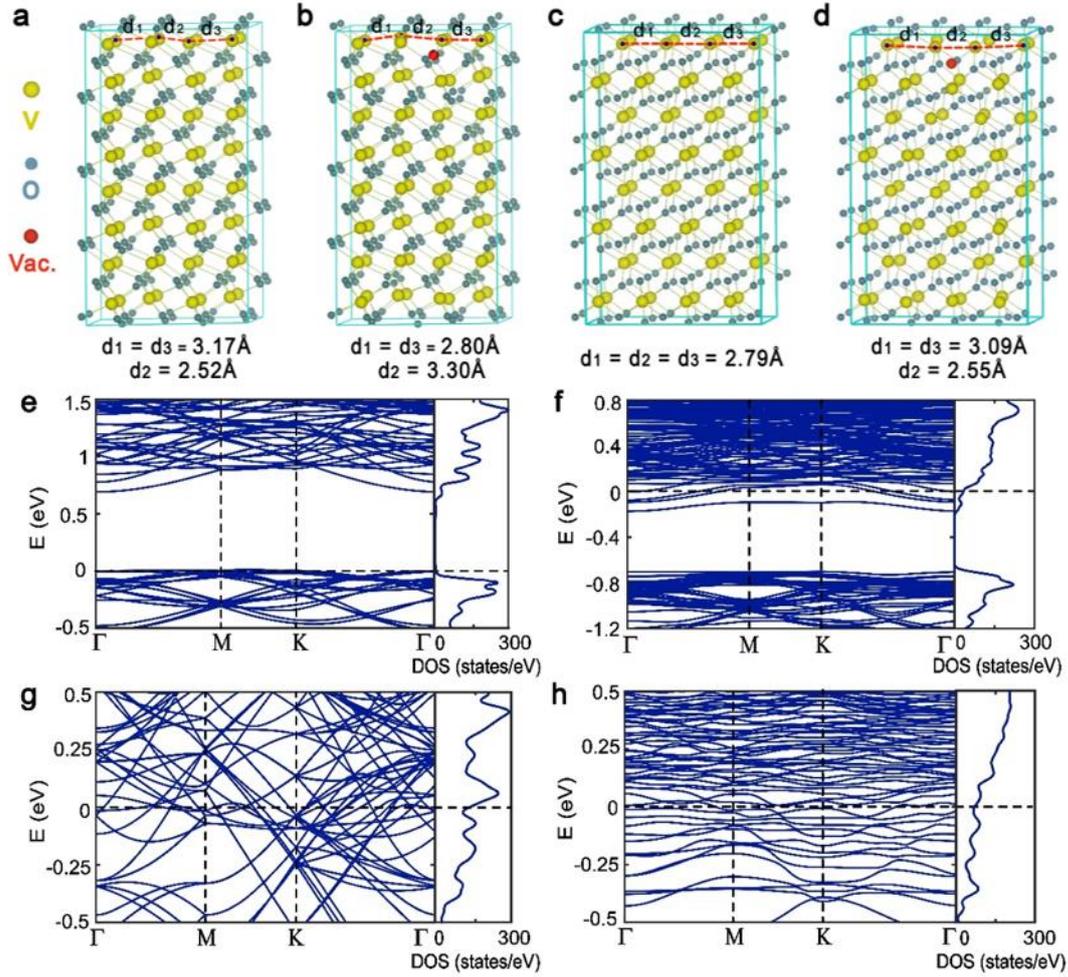

**Figure 4. The calculated atomic and electronic structures of VO$_2$ in two phases, before and after introducing an oxygen vacancy.** The superlattice structures include the ones of (**a**) the pristine *M*-phase, (**b**) the *M*-phase with one oxygen vacancy, (**c**) the pristine *R*-phase, and (**d**) the *R*-phase with one oxygen vacancy. In (**a**)-(**d**), the yellow, blue and red balls denote the vanadium atom, the oxygen atom and the oxygen vacancy, respectively, and the representative V-V bonds are shown by red dashed lines. The corresponding electronic structures of the four superlattices are shown in (**e**)-(**h**), with the band structure and density of states in each panel. The Fermi level is set to zero energy and donated by horizontal dashed lines in (**e**)-(**h**).